\documentclass[	%preprint,%
						superscriptaddress,%
						twocolumn,%
						a4paper,%			Format A4
					]{revtex4}

\usepackage[frenchb,english]{babel}
\usepackage[T1]{fontenc}
\usepackage{upgreek}
\usepackage{textcomp} 				% pour avoir de jolis symboles degré
\usepackage{graphicx}			% insert .eps ([dvips] aus Mael's Version entfernt, um die Bilder zu sehen)
\usepackage{epstopdf}				% einfügen von .eps mit pdflatex  
\usepackage{color}					% def de la page
\usepackage[nice]{nicefrac} 		% pour avoir des fractions 1/2 jolies
\usepackage{amssymb,amsmath}		% polices et symboles math utiles pour eqnarray et subequations
\usepackage{array}					% pour des tableaux évolués
\usepackage{ulem}
\definecolor{vert}{rgb}{0.5,0.758,0.5}
\definecolor{bleufonce}{rgb}{0,0,0.516}
\definecolor{orange}{rgb}{1,0.516,0}

\begin{document}

\title{Temperature evolution of the band-gap in BiFeO$_3$ traced by resonant Raman scattering}
\date{\today}
\author{Mads Christof Weber}
\affiliation{Materials Research and Technology Department, Luxembourg Institute of Science and Technology, 41 rue du Brill, L-4422 Belvaux, Luxembourg}
\affiliation{Physics and Materials Science Research Unit, University of Luxembourg, 41 Rue du Brill, L-4422 Belvaux, Luxembourg}
\author{Mael Guennou}
\affiliation{Materials Research and Technology Department, Luxembourg Institute of Science and Technology, 41 rue du Brill, L-4422 Belvaux, Luxembourg}
\email{mael.guennou@list.lu}
\author{Constance Toulouse}
\affiliation{Laboratoire Mat\'eriaux et Ph\'enom\`enes Quantiques (UMR 7162 CNRS), Universit\'e Paris Diderot--Paris 7, 75205 Paris cedex 13, France}
\author{Maximilien Cazayous}
\affiliation{Laboratoire Mat\'eriaux et Ph\'enom\`enes Quantiques (UMR 7162 CNRS), Universit\'e Paris Diderot--Paris 7, 75205 Paris cedex 13, France}
\author{Yannick Gillet}
\affiliation{Universit\'e catholique de Louvain, Institute of Condensed Matter and Nanosciences, Nanoscopic Physics, Chemin des \'Etoiles 8, bte L7.03.01, 1348 Louvain-la-Neuve, Belgium}
\author{Xavier Gonze}
\affiliation{Universit\'e catholique de Louvain, Institute of Condensed Matter and Nanosciences, Nanoscopic Physics, Chemin des \'Etoiles 8, bte L7.03.01, 1348 Louvain-la-Neuve, Belgium}
\author{Jens Kreisel}
\affiliation{Materials Research and Technology Department, Luxembourg Institute of Science and Technology, 41 rue du Brill, L-4422 Belvaux, Luxembourg}
\affiliation{Physics and Materials Science Research Unit, University of Luxembourg, 41 Rue du Brill, L-4422 Belvaux, Luxembourg}
\email{jens.kreisel@list.lu}

\begin{abstract}
Knowledge of the electronic band structure of multiferroic oxides, crucial for the understanding and tuning of photo-induced effects, remains very limited even in the model and thoroughly studied BiFeO$_3$. Here, we investigate the electronic band structure of BiFeO$_3$ using Raman scattering with twelve different excitation wavelengths ranging from the blue to the near infrared. We show that resonant Raman signatures can be assigned to direct and indirect electronic transitions, as well as in-gap electronic levels, most likely associated to oxygen vacancies. Their temperature evolution establishes that the remarkable and intriguing variation of the optical band-gap can be related to the shrinking of an indirect electronic band-gap, while the energies for direct electronic transitions remains nearly temperature independent.
\end{abstract}


\maketitle

The design of multifunctional materials is one of the most investigated topics in current condensed matter physics. Materials possessing more than one ferroic order -- multiferroics -- have attracted a particular interest, as they can intrinsically combine magnetic, ferroelectric and elastic properties. More recently, the interest for multiferroics has also included their interactions with light, with perspectives for a variety of possible applications \cite{Eerenstein2006,Bibes2012,Ramesh2007}. Most of the recent stimulating work on materials with both photo- and ferroelectric properties - called "photoferroelectrics" \cite{Fridkin1979} - has focused on $AB$O$_3$ perovskites, and specifically on BiFeO$_3$ for its above-band gap photovoltages, anomalous photovoltaic effects, photo-conductive domain walls etc. \cite{Yang2010,Alexe2011,Seidel2011,Kundys2010,Bhatnagar2013}. More recently, the crucial role of in-gap electronic states for the photovoltaic effect has been demonstrated \cite{Yang2015}. Yet, the electronic structure in BiFeO$_3$ in general and its band gap in particular, in spite of their importance for photoferroelectric effects, remain rather poorly characterized experimentally, and theoretical works remain scarce. According to absorption studies on single crystals \cite{Xu2009} and thin films \cite{Basu2008,Ihlefeld2008,Kumar2008,Hauser2008}, the optical band gap at ambient conditions lies at approximately 2.7~eV. It is also experimentally established that the optical band gap of BiFeO$_3$ shrinks with increasing temperature down to 1.3~eV at 1200~K where it then closes abruptly as a consequence of an insulator-to-metal phase transition concomitant with a structural transition to the so-called $\upgamma$-phase \cite{Palai2008}. Both this rapid shrinking of the optical band-gap, three times steeper than in BaTiO$_3$ \cite{Wemple1970}, and the insulator-to-metal transition set BiFeO$_3$ apart from conventional non-magnetic ferroelectrics. It is not yet understood how this gradual shrinking of the optical band gap relates to the electronic structure. One of the reasons for the lack of knowledge is that the experimental investigation of band-to-band transitions is difficult to address experimentally in ferroelectric and multiferroic oxides because the absorption onset is broad, especially when compared to the generally sharp transitions in classical semiconductors. The appearance of Urbach tails, notably at higher temperatures, complicates the quantitative analysis and the discrimination of direct and indirect transitions. Other classical techniques are rapidly limited by thermal effects, charging of the insulating samples (ARPES) or require synchrotron sources (RIXS). In this letter, we make use of resonant Raman scattering, which occurs when the incident or scattered photon energy is close to an electronic transition. These resonant effects have been investigated in many details in semiconductors \cite{Leite1969,Scott1969a,Scott1970a,Hayes1978} but have been much less investigated in ferroic perovskites; they have been used either as a probe for the structural behavior \cite{Rouquette2006,Flor2014,Fraysse2012} or electronic transitions involving mostly charge transfer phenomena \cite{Fujioka2004,Andreasson2007,Andreasson2008,Andreasson2009} that are difficult to transfer to the case of pure BiFeO$_3$.

\begin{figure*}[t]
\begin{center}
\includegraphics[width=0.9\textwidth]{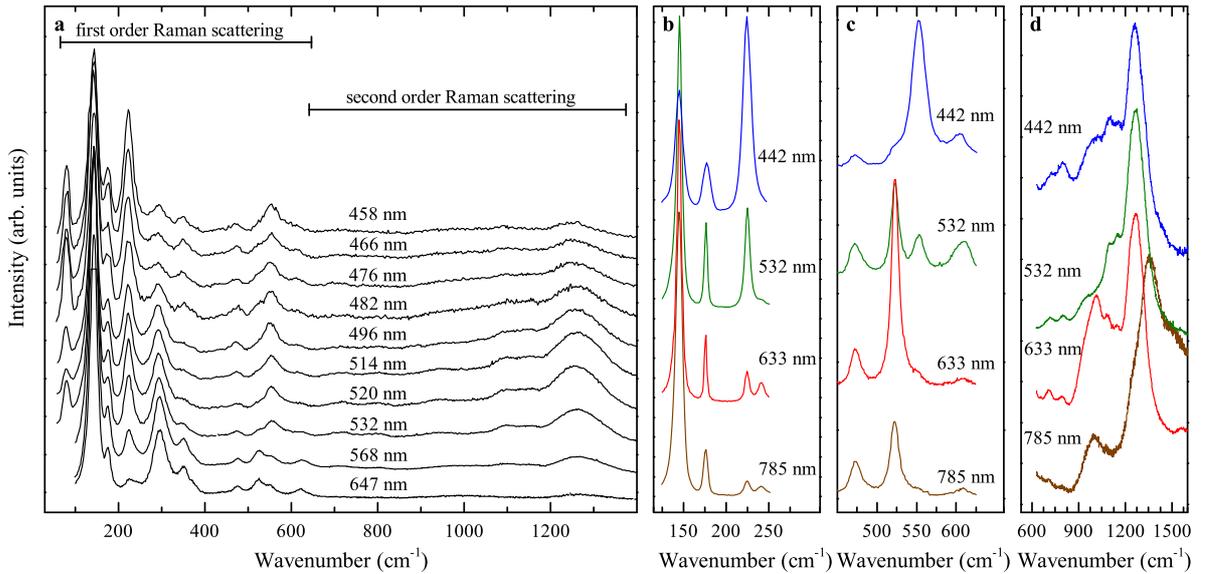}
\caption{a) Raman spectra for different excitation wavelengths ranging from 458~nm to 647~nm. The spectra were recorded at room temperature. For better visibility, the spectra were normalized to the integrated intensity of the first order Raman scattering process. b) and c) Zoom into two low- and high-wavenumber regions. For a better definition of the Raman bands, the spectra were measured at 93~K. d) Second order Raman spectra for several excitation wavelengths at ambient condition.}
\label{fig:spectra}
\end{center}
\end{figure*}

In a first step we have investigated $[001]_\mathrm{pc}$-oriented BiFeO$_3$ single crystals at room temperature with twelve different excitation wavelengths ranging from 442~nm to 785~nm. Samples were grown by the flux method as described elsewhere \cite{Haumont2008,Lebeugle2007}. Measurements were performed independently on two different setups in backscattering configuration (an inVia Renishaw Raman spectrometer in micro-Raman mode and a Jobin Yvon T64000 in a macro-Raman mode). In order to take into account the effect of the dependence of the laser wavelength on spectral response, the spectra from the Renishaw inVia spectrometer were normalized with respect to a reference CaF$_2$ crystal \cite{Cardona1982}. The spectral response of the T64000 was calibrated using a white lamp and the spectra were corrected for the $\omega^4$ dependence \cite{Cardona1982}. In both cases the wavelength-dependent absorption in BiFeO$_3$ was corrected using the data from Ref.~\cite{Xu2009}. The Raman spectra were corrected for thermal occupation by $(n+1)$ and $(n+1)^2$ for first and second order scattering, respectively, where $n$ is the Bose-Einstein occupation number. Great care was taken to exclude any spurious effect related to crystal orientation, laser polarization, or oblique modes \cite{Hlinka2011}, i.e. the orientation of the analyzed domain with respect to the direction of propagation and polarization of the laser was carefully checked in all cases.  Fig.~\ref{fig:spectra} (a) shows the normalized Raman spectra, which are well defined with sharp bands, especially at lower wavenumbers, and with Raman frequencies in agreement with reported literature data (typically measured with red or green exciting lasers) \cite{Fukumura2007,Cazayous2007,Palai2010,Hlinka2011}. The most prominent feature of the spectra is that the relative intensity of the different Raman bands greatly depends on the wavelength used. Figures~\ref{fig:spectra} (b) and (c) make this more apparent by magnification of two narrow spectral regions: for example, the band at 230~cm$^{-1}$ is very strong at 442~nm but hardly observable at 785~nm; similar changes are observed for almost all modes when carefully considering the full series. The intensity of the second-order Raman spectrum, visible as broad bands in the 1000--1400~cm$^{-1}$ range (see Fig.~\ref{fig:spectra} (d)), also depends on the exciting wavelength in agreement with previous reports \cite{Cazayous2009}. 

Variations of the total intensity and intensity ratios between different bands are characteristic features when passing from non-resonant to resonant Raman scattering, i.e. when the energy of the incident light approaches that of an electronic transition. More precisely, resonant Raman scattering is observable only when the Raman scattering process with a particular electronic transition is symmetry-allowed. This situation leads to a divergence of the Raman tensor and therefore a strong enhancement in the Raman scattering signal, where the scattering intensities of the Raman modes are then dominated by the transition probabilities associated to this particular electronic transitions between real -- and no longer virtual -- electronic states~\cite{Hayes1978,YuCardona}. Because the transition probability differs for each vibrational distortion -- and each Raman mode accordingly, the intensity ratios between different modes change with respect to non-resonant situations. In general, not only intrinsic but also defect levels can be involved in this process. Importantly, because first-order Raman scattering is a one-phonon process, only direct electronic transitions can fulfill momentum conservation. In contrast, second-order Raman scattering involves a combination of phonons anywhere in the Brillouin zone, so that second-order resonance processes may involve both direct and indirect interband electronic transitions. 

\begin{figure}[t]
\begin{center}
\includegraphics[width=0.5\textwidth]{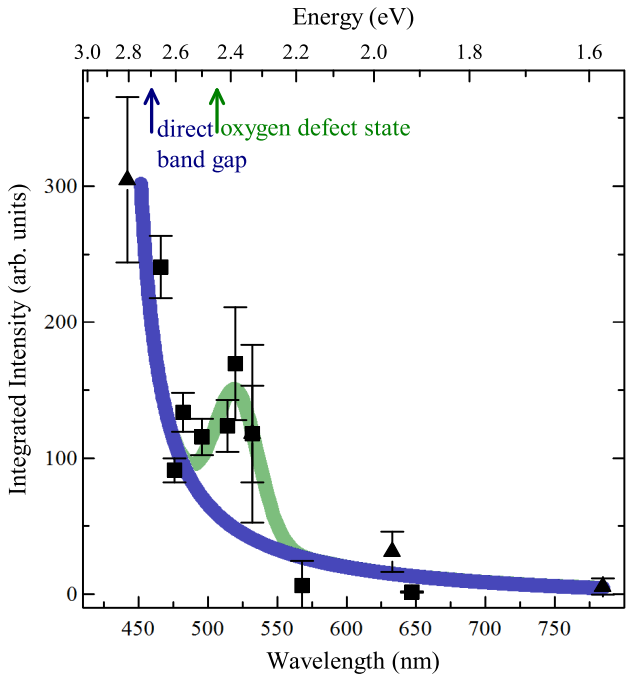}
\caption{Evolution of the integrated intensity of the first order spectra from 100~cm$^{-1}$ to 650~cm$^{-1}$ at room temperature. The triangles and squares are spectral data measured in separate runs and with different spectrometers. Appropriate spectrometer and absorption corrections have been applied as explained. The integrated intensity of the 647~nm spectrum (which represents the lowest intensity) has been set to one; other spectra where normalized accordingly, to provide a measure for the enhancement. The solid lines are guides for the eye.}
\label{fig:resonance}
\end{center}
\end{figure}

In spite of the variations in intensity ratios, all the bands in the spectra of Fig.~\ref{fig:spectra} remain visible, unlike classical cases observed for semiconductors where LO modes contribute to Fröhlich interactions and are selectively enhanced \cite{YuCardona}. All bands in BiFeO$_3$ can therefore be considered resonance-Raman-active. The resonance enhancement can then be quantified by integrating the intensity over the full first-order spectral range (i.e. up to 650~cm$^{-1}$). This integrated intensity is plotted in Fig.~\ref{fig:resonance} for all wavelengths, revealing intensity changes by up to two orders of magnitude across the series. It remains constant between 785 and 568~nm but then increases sharply when approaching the green laser lines leading to a first maximum at around 520~nm. Towards higher energies (smaller laser wavelengths), the integrated intensity decreases slightly before a very strong second enhancement occurs under excitation with blue laser light (465~nm or 2.67~eV). This strong increase correlates with the reported value of 2.7~eV for the band-gap. The intensity maximum at approximately 520~nm (2.38~eV) shall then correspond to a resonance process involving in-gap electronic states. Defect states related to oxygen vacancies are obvious candidates and, indeed, have been reported experimentally at 2.45~eV \cite{Hauser2008}, which is also consistent with the 2.2~eV value predicted by first-principles calculations \cite{Clark2009}. We therefore assign this second resonance enhancement at 520~nm with a valence band to defect transition involving oxygen vacancies.

When comparing the Raman spectra at different excitation wavelengths, one should bear in mind that the absorption by the sample is also wavelength dependent, so that a change in the exciting laser wavelength leads to a change in penetration depth. In materials having a "skin layer" with different structural properties from the bulk, such as ferroelectrics in general and BiFeO$_3$ in particular \cite{Marti2011}, this effect can in principle affect the very shape of the Raman signature. However, this effect can be safely ruled out in our study considering that: (i) a structural distortion would be expected to cause shifts of the Raman frequencies, which are not observed and (ii) the skin layer in BiFeO$_3$ is not thicker than 10~nm \cite{Marti2011,Domingo2013}, whereas changes of the Raman spectrum are visible for lasers penetrating much deeper in the bulk: applying Beer-Lambert's law with the absorption data of Xu et al. \cite{Xu2009} gives several micrometers for near-infrared to red light, several hundreds of nanometers for green light (532~nm) and 60--80~nm for blue light (442~nm). 

\begin{figure}[t]
\begin{center}
\includegraphics[width=0.5\textwidth]{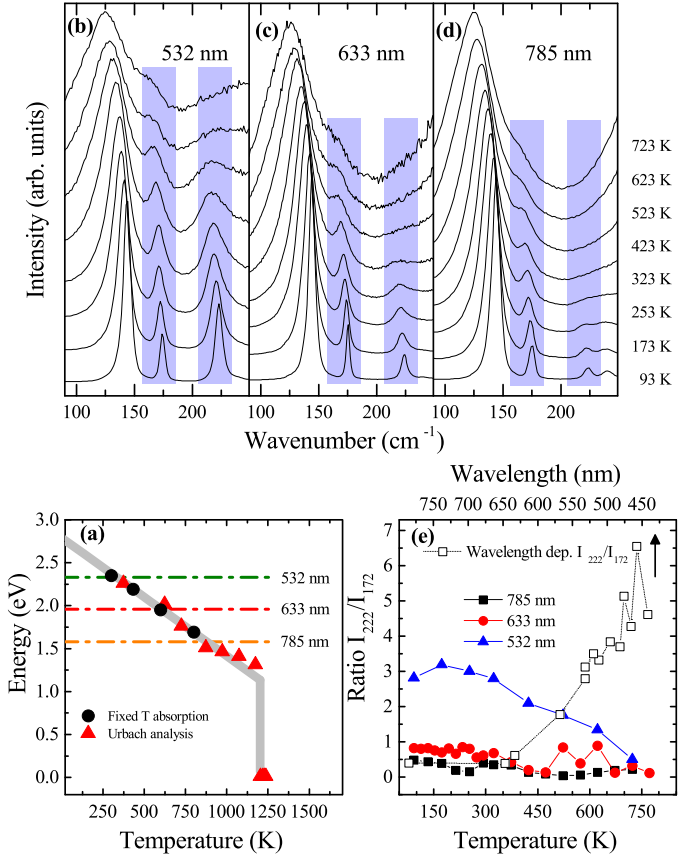}
\caption{(a) Band gap values versus temperature taken from Ref.~\cite{Palai2008}, together with the energies of the 532~nm, 633~nm and 785~nm laser lines. (b)--(d) Temperature-dependent Raman spectra in the low-frequency region for the 532~nm, 633~nm and 785~nm laser lines. The regions in blue highlight two bands that show a significant change in their intensity ratio when crossing resonance (see Figure \ref{fig:spectra} (b)). (e) Evolution of the intensity ratio of the bands located around 172 and 223~cm$^{-1}$ (highlighted in (b)-(d)) with temperature (bottom axis), and with excitation energy at ambient temperature (top axis).}
\label{fig:spectratemp}
\end{center}
\end{figure}

Because changes of the band gaps imply changes in the Raman resonance conditions, resonant Raman spectroscopy can now be used to track its temperature evolution, with the objective to elucidate the intriguing shrinking of the optical gap at high temperatures. This principle is illustrated in Figure \ref{fig:spectratemp} (a), where the band gap values determined by Palai et al. \cite {Palai2008} are displayed together with the energies of three selected laser lines used in this study. Strong changes in the resonance conditions are expected when the laser energy is close to the reported band gap, which should be reflected in the intensity and the shape of the Raman signature. 

Figure \ref{fig:spectratemp} (b)--(d) shows the low-frequency part of the first-order Raman spectra for temperatures ranging from 93~K to 773~K and for laser excitation wavelengths 532~nm, 633~nm and 785~nm, i.e. for energies that coincide with the reported band-gap at different temperatures. In all cases, the temperature evolution of the Raman spectra follows a typical behaviour characterized by thermal broadening and a generalized low-frequency shift of all bands with increasing temperature. The intensity ratios between modes do not change significantly compared to the intensity ratio changes observed for different excitation wavelength (see Fig.~\ref{fig:spectra}). This is quantified in Fig.~\ref{fig:spectratemp} (e) that shows the evolution of the intensity ratio for the two highlighted vibrational bands as an example. This ratio does not show any sign of the dramatic increase that characterizes the resonant regime that would be expected at high temperatures when the laser energy meets the reported optical band gap. This leads to the at first sight surprising conclusion that the resonance conditions at a given wavelength do not change with temperature. Considering that only direct electronic transitions can be involved in first-order Raman resonances, we come to the following conclusions: (i) the temperature evolution of the optical band-gap cannot be related to a shrinking of a direct electronic band-gap and (ii) the electronic levels underlying the resonance Raman effect under illumination with 442~nm (assigned to direct electronic transitions) and 532~nm (oxygen vacancies) are nearly temperature-independent.

\begin{figure}[ht]
\begin{center}
\includegraphics[width=0.5\textwidth]{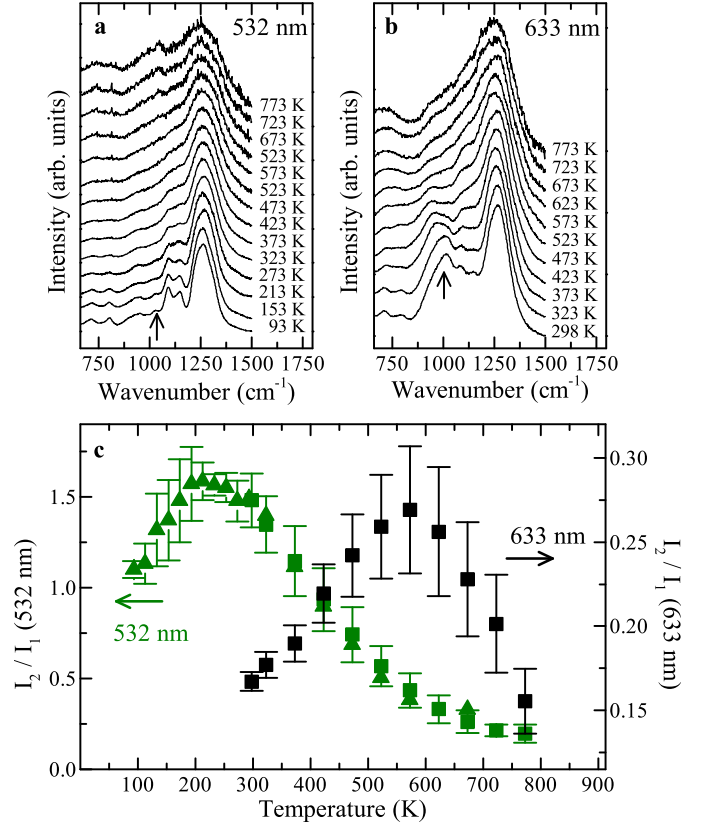}
\caption{a) and b) Temperature-dependent second order Raman spectra under excitation of 633~nm and 532~nm laser light. c) Temperature-dependent ratio of the integrated intensities of the second ($I_2$) to first order ($I_1$) Raman scattering spectra.}
\label{fig:2ndorder}
\end{center}
\end{figure} 

In a second step, we have scrutinized the second-order Raman spectra. Figures~\ref{fig:2ndorder} (a) and (b) present the temperature evolution of the second-order Raman spectra for excitation of 532~nm and 633~nm. The figures show a strong qualitative change in the intensity signature, which is particularly apparent when comparing the relative intensities of the broad peaks at 1010 and 1260~cm$^{-1}$ under excitation by the 633~nm laser line. The band at 1010~cm$^{-1}$ is strongly reduced at higher temperatures and the shape of the spectrum gradually becomes similar to the spectrum obtained for 532~nm laser at ambient temperature, albeit with differences due to different thermal broadenings. Conversely, the 532~nm laser line spectra at low temperature revealed a small but distinct band at 1010~cm$^{-1}$ (indicated by the arrow in Figure~\ref{fig:2ndorder} (a) and (b)). These qualitative changes were accompanied by a change in the overall intensity, which we quantify by the ratio between the integrated intensity ($I_2$) of the second-order spectrum divided by the intensity of the first-order spectrum ($I_1$). For both excitation wavelengths, $I_2/I_1$ exhibited a maximum in their temperature dependence. The observed maxima at 573~K for 633~nm and at 223~K for 532~nm retrace the temperature dependence of the reported optical band gap \cite{Palai2008}. These changes in shape an intensity of the second-order Raman signature contrasts with the temperature-independent shape of the first-order spectrum. Because any strong change in the energies of direct electronic transitions was ruled out, the observed maxima of the ratio $I_2/I_1$ can be related to a temperature evolution of the indirect gap. Other possible phenonema are more difficult to reconcile with the experimental observations. Notably, the Franck-condon mechanism explaining the resonance behavior reported in past studies \cite{Fujioka2004,Andreasson2008} does not have the same signature, both at ambient condition (e.g. no 3$^\mathrm{rd}$-order scattering is observed in our studies) and as a function of temperature. Also, a magnetism-related change in the electron-phonon coupling could be admittedly hypothesized from the observation that the maximum for $I_2/I_1$ at 633~nm lies in the vicinity of $T_N$, but it is not clear then how to explain the wavelength dependence of this maximum, and the fact that no such anomaly is observed in the vicinity of $T_N$ in the related compound LaFe$_{0.96}$Cr$_{0.04}$O$_3$ \cite{Andreasson2008}.

Putting together our experimental observations therefore leads to the conclusion that the shrinking of the optical band-gap is related to a temperature-evolution of an electronic band-gap with indirect character. This supports the view that BiFeO$_3$ is an indirect semiconductor, at least above room temperature. On the other hand, the energies of direct electronic transitions show no temperature dependence, which implies in turn non-trivial temperature modifications of the electronic band structure. Even though only hypothesis can be formulated at this stage for the details of this temperature dependence, we note that the hypothesis of a strong electron-phonon coupling inducing an important renormalization of the electronic levels is consistent with the fact that the highest valence band originates predominantly from the oxygen p-states. The low relative mass of the oxygen atom then leads naturally to comparatively large nuclear displacements in the corresponding phonon modes, and hence the largest temperature dependence of oxygen-based bands. Starting from valence and conduction bands considered as nearly flat \cite{Palai2008,Shelke2012a}, an increased curvature of the bands at high temperature can explain the different behavior of the direct and indirect gaps. Further work will be needed for confirmation of this picture, including both calculations at finite temperatures and experiments to follow carefully the optical absorption with temperature and probe directly electronic levels. More generally, resonant Raman scattering appears as a powerful tool to follow electronic excitations up to relatively high temperatures, in a way that should be applicable beyond BiFeO$_3$ to a broad range of materials showing absorption in the visible range, and thereby to contribute to the understanding of electronic structures in photoferroelectrics.

\begin{acknowledgments}
The authors thank J. F. Scott (Cambridge, UK) and S. Siebentritt (University of Luxembourg) for useful discussions as well as R. Haumont (Université Paris Sud), and D. Colson for providing high-quality single crystals. The authors acknowledge financial support from the Fond National de Recherche Luxembourg through a FNR-PEARL award (FNR/P12/4853155/Kreisel). Y.G. acknowledges financial support from the Fonds National de la Recherche Scientifique through a FNRS fellowship.
\end{acknowledgments}

%%%%%%%%%%%%%%%%%%%%%%%%%%%%%%%%%%%%%%%%%%%%%%%%%%%%%%%%%%
%%%%%%%%%% Bibliographie
%%%%%%%%%%%%%%%%%%%%%%%%%%%%%%%%%%%%%%%%%%%%%%%%%%%%%%%%%%
%\bibliographystyle{aip}
%\bibliography{biblio}

\end{document}